\documentclass[conference]{IEEEtran}
\IEEEoverridecommandlockouts
\usepackage{cite}
\usepackage{amsmath,amssymb,amsfonts}
\usepackage{algorithm}
\usepackage{algorithmic}
\usepackage{graphicx}
\usepackage{textcomp}
\usepackage{xcolor}
\usepackage{float}
\usepackage{caption}
\usepackage{subcaption}
\usepackage{array}
\usepackage{tabularx}
\usepackage{multirow}
\usepackage{lipsum}
\usepackage{booktabs}
\usepackage{makecell}

\def\BibTeX{{\rm B\kern-.05em{\sc i\kern-.025em b}\kern-.08em
    T\kern-.1667em\lower.7ex\hbox{E}\kern-.125emX}}
\begin{document}

\title{Proactive and Automatic Underfrequency Load Shedding via PMUs and Particle Filters  \\
\thanks{This work was partially funded by the U.S. Department of Energy
under Contract DE-AC05-76RL01830.}
}

\author{\IEEEauthorblockN{Gian Paramo}
\IEEEauthorblockA{\textit{Electrical and Computer Engineering} \\
\textit{University of Florida}\\
Gainesville, FL, USA \\
gparamo@ufl.edu}
\and
\IEEEauthorblockN{Arturo Bretas}
\IEEEauthorblockA{\textit{Distributed Systems Group} \\
\textit{Pacifc Northwest National Laboratory}\\
Richland, WA, USA \\
arturo.bretas@pnnl.gov}
\and
\IEEEauthorblockN{Newton Bretas}
\IEEEauthorblockA{\textit{Electrical and Computer Engineering} \\
\textit{University of Sao Paulo}\\
Sao Carlos, SP, Brazil \\
ngbretas@sc.usp.br}
}
\IEEEoverridecommandlockouts\IEEEpubid{\makebox[\columnwidth]{978-1-6654-8537-1/22/\$31.00~\copyright~2022 IEEE \hfill} \hspace{\columnsep}\makebox[\columnwidth]{ }}

\maketitle

\begin{abstract}
Underfrequency (UF) load shedding schemes are traditionally implemented in two ways: One approach is based on manual load shedding, with system operators requesting loads to be shed ahead of anticipated stressful operating conditions.
Manual load shedding is usually done through phone calls. The second method is automatic load shedding via underfrequency relays. Using static settings, these schemes can be designed to operate in stages and drop previously identified loads. The main limitation of traditional load shedding schemes is that they are reactive and leave little room for optimized corrective actions. This work presents a proactive and automatic underfrequency load shedding solution for power systems.
Measurements are captured via phasor measurement units (PMUs) at relatively low sampling rates of 30 Hz. These measurements are then processed by particle filters who predict the future state of the system's frequency. Based on these predictions excess load is determined and shed. Comparative case studies are performed in simulated environments. Easy-to-implement models, without hard-to-derive parameters, highlight potential aspects for real-life implementation.    
\end{abstract}

\begin{IEEEkeywords}
underfrequency load shedding, particle filters, phasor measurement units, power system protection. 
\end{IEEEkeywords}

\section{Introduction}

Currently, the majority of automatic underfrequency load shedding solutions are built on decentralized architectures, and operate at the feeder level. UF relays are installed at feeders considered non-critical by power system operators \cite{relaybook}. Once the frequency seen by the relay drops below a threshold, the relay issues a trip signal and the entire circuit is disconnected. 
Small degrees of selectivity and coordination can be achieved by applying time delays or multiple pick-up settings \cite{Sig}.
Despite these improvements, these solutions, referred to in this work as traditional UF schemes, take a drastic and unforgiving approach: if the relay sees the system's frequency drop, the customers on that circuit are disconnected. 
Traditional UF schemes are reactive, as corrective actions occur only after a disturbance has been observed. UF load shedding strategies of this type suffer from two commonly observed issues: delayed response and overshedding \cite{Sig}. 

While traditional UF schemes are by far the most widely utilized solution, more intricate techniques leveraging the processing power of digital relays have been suggested \cite{relaybook}. Semi-adaptive techniques, those considering time-derivatives, and fully adaptive techniques 
have been proposed \cite{ander, pasand, abdel}. While these offer improved performance compared to traditional schemes, the problems of delayed response and overshedding are still present. Details regarding frequency stability and mitigation methods can be found in \cite{kundur94, kundur04}. 

Clearly, corrective actions in the face of UF events could be orchestrated better; however, a well choreographed response requires time. Precious seconds can be gained by switching from the traditional reactive approach into a proactive scheme \cite{GP}.  


Predictive schemes are not exactly new, one such approach was suggested in \cite{bretas} over two decades ago. 
The method presented in \cite{bretas} has the same foundation as many of the techniques proposed today: Measurements are collected via PMUs, data is processed as a time series, and short term predictions are made. 
An approach similar to \cite{bretas} is presented in \cite{larsson}. However, \cite{larsson} focuses on steady state UF mitigation. 
This technique makes decisions based on the predicted steady state value of frequency after a disturbance is detected.  
More recently, an elegant solution was presented in \cite{rudez}. Measurements taken via PMUs are processed by a prediction algorithm based on simple polynomial curve fitting. 

One significant limitation of the techniques found in literature is that they cannot be supported by technology that is currently available or already in service. For instance, these approaches normally rely on PMU measurements collected at high sampling rates (in some cases over 100 Hz as in \cite{bretas}), meanwhile, the PMU sampling rate of modern digital relays is only 30 Hz\cite{sel421}. This is the equipment demographic this method aims to exploit.

Another limitation found in contemporary UF frameworks is the use of non-adaptive physics-based models. Any model carries an implicit modelling uncertainty, and this problem is compounded when the state of the physical system changes. 

In light of the limitations of traditional and contemporary UF solutions, this work presents a proactive framework for automatic UF control where model uncertainty is continuously updated through particle filters \cite{Elf}.
This work found that particle filters offer an exciting degree of accuracy and robustness with relatively modest requirements in terms of hardware and expertise. 

A different approach to UF mitigation is introduced in \cite{pul}. Deviations in frequency are compensated in real-time by actuating DERs. 
The results of \cite{pul} are promising and are used as a benchmark in the third case study presented in this work.

The rest of the paper is structured as follows: Section II provides a mathematical introduction into the particle filter and the equations used in system modeling.
Section III showcases three case studies and discusses the findings. Concluding remarks are given in Section IV.   

\section{Theoretical Background}

\subsection{Particle Filter}
\begin{figure}
\centering
\captionsetup{justification=centering}
\includegraphics[width=7.5cm]{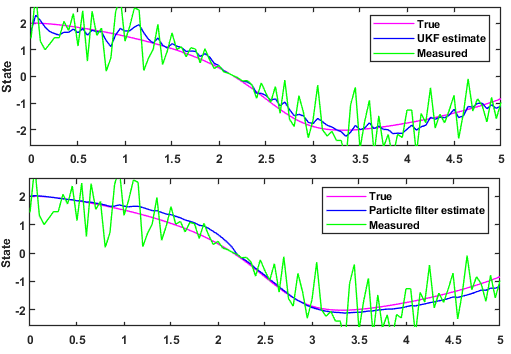}
\caption{UKF estimate (top) vs PF estimate (bottom).\label{pic00}}
\end{figure}

In the last decade filtering techniques, such as the Kalman filter (KF) have gained attention from researchers in the area of power systems. 
This has been driven by advancements in hardware, computing power, and the need to establish a framework for dynamic state estimation \cite{Junbo}. 
While KF based techniques have provided encouraging results, some limitations of the KF have been observed. 
In particular, a common assumption that data points follow a Gaussian distribution has been called into question in \cite{wang}. 
A lesser known filtering technique, the particle filter (PF), solves this limitation by making multiple predictions for each state being tracked, with each prediction having a different probability. 
At each time step predictions and corrections, equivalent to those in the KF are performed considering data from previous time steps, an underlying system model, and the predictions made. 
The result is a filter that is more flexible than the KF as illustrated in Figure \ref{pic00}.   

When working with the PF, the optimal solution in Bayesian form is calculated as a sum of samples. These samples are referred to as particles, and each one is assigned a weight. 
\begin{equation}
p(x_{0:k}|z_{1:k}) \approx \Sigma^{N_{s}}_{i=1}w^{i}_{k}\delta(x_{0:k}-x^{i}_{0:k})
\end{equation}
A set containing $N_{s}$ samples and weights can be expressed as $\{w^{i}_{k}, x^{i}_{0:k}\}^{N_{s}}_{i=1}$. The weight, or importance of each sample $x^{i}_{0:k}$ is represented by $w^{i}_{k}$. The sum of all weights $\Sigma^{N_{s}}_{i=1}w^{i}_{k} = 1$. Samples thought to be of higher accuracy are given a higher weight relative to samples of lower accuracy. Finally, the Dirac delta function is represented by $\delta(\cdot)$. Weights are computed via (2):
\begin{equation}
w^{i}_{k} \: \alpha \: w^{i}_{k-1}\frac{p(z_{k}|x^{i}_{k})p(x^{i}_{k}|x^{i}_{k-1})}{q(x^{i}_{k}|x^{i}_{k-1},z_{k})}
\end{equation}
Without knowledge of a posterior, but given density $q$, the relationships in (2) can be used to assign particle weights. This equation can be thought of as a ratio between posterior and importance density for each particle. With further manipulations it can be shown that the corresponding posterior can be expressed as:
\begin{equation}
p(x_{k}|z_{1:k}) \approx \Sigma^{N_{s}}_{i=1}w^{i}_{k}\delta(x_{k}-x^{i}_{k})
\end{equation}
A key takeaway from (3) is that as the number of particles is increased, the solution moves closer to the real values. 

Resampling in the PF plays a similar role to the correction step in the KF. Resampling attempts to correct imbalances in weight assignments that might skew the overall performance of the filter.

 \begin{algorithm}[H]
 \caption{Particle filter with resampling}
 \begin{algorithmic}[1]
 \renewcommand{\algorithmicrequire}{\textbf{Input}}
 \renewcommand{\algorithmicensure}{\textbf{Output}}
 \REQUIRE $\{x^{i}_{k-1},w^{i}_{k-1}\}^{N_{s}}_{i=1}, z_{k}$
 \ENSURE  $\{x^{i}_{k},w^{i}_{k}\}^{N_{s}}_{i=1}$ \\
  \textit{$w_{sum = 0}$} \\
  \FOR {$i = 1,...,N_{s}$} 
  \STATE draw sample $x^{i}_{k} \approx q(x^{i}_{k}|x^{i}_{k-1},z_{k})$ \\
  assign weight $w^{i}_{k}$ using (2)\\
  $w_{sum} = w_{sum} + w^{i}_{k}$ \\
  \ENDFOR \\
  \FOR {$i = 1,...,N_{s}$} 
  \STATE $w^{i}_{k} = w^{i}/w_{sum}$ 
  \ENDFOR \\
  \STATE Resample $N_{s}$ particles with replacement \\
  \FOR {$i = 1,...,N_{s}$}  
  \STATE $w^{i}_{k} = 1/N_{s}$ 
  \ENDFOR
 \end{algorithmic}
 \end{algorithm}

A complete derivation of the PF was omitted due space constraints but it can be found in \cite{Elf}. A comparison among several types of Bayesian filters can be found in \cite{Arul}.  

The PF can be considered a non-Gaussian extension of the KF. The trade-off for this increase in flexibility is a modest increase in complexity. For this reason, conservative processing delays are included in the case studies presented in Section III. 

\subsection{System Frequency}
A combination of hard and soft thresholds were use in this work. The hard thresholds are those seen in traditional UF schemes. Action is taken after frequency drops below a set point. For the soft thresholds, actions are taken based on the rate of change of frequency:    

\begin{equation}
R = \frac{f_{2}-f_{1}}{dt}\label{eq4}
\end{equation}

were $R$ is the average rate of change in frequency, $f_{1}$ is the initial frequency, while $f_{2}$ is the frequency at the end of the time window $dt$. These soft thresholds have delays corresponding to the magnitude of the difference between the two frequency measurements. These values are presented as a lookup table in \cite{sel751}, with large values of $R$ (2.33 to 15 Hz/sec) having a delay of only 3 cycles, while smaller values of $R$ (0.33 to 0.37 Hz/sec) have a corresponding delay as large as 21 cycles.   

\subsection{Prediction Problem}
In their original formulation, the KF and the PF make predictions one time step into the future ($k+1$). A simple way to extend the horizon of these predictions is to feed artificial data points (ADPs) into the filter. The number of ADPs required can be found by considering the sampling frequency and the number of seconds into the future one wishes to predict:
\begin{equation}
N_{ADP} = \frac{t_{p}}{f_{s}}\label{eq11}
\end{equation}
Here $N_{ADP}$ is the number of artificial measurements required, $t_{p}$ is the number of seconds into future, and $f_{s}$ is the sampling frequency of the measuring device.
In order to emulate the dynamic nature of the system, the first and second deviates are calculated for each of the last ten time steps, corresponding to the last ten data points before a prediction is made. The derivatives are then averaged, before being used to systematically adjust the last data point received to produce a vector of ADPs. This is a sequential process based on the following equation:
\begin{equation} \label{eq1}
\begin{split}
ADP_{i} & = ADP_{i-1} + t_{s}f' + t^2_{s}f'' \\
\end{split}
\end{equation}
Where $ADP_{i-1}$, is the previous ADP. The average first derivative is represented by $f'$, while the average second derivative is represented by $f''$. Finally, $t_{s}$ represents the time window of the derivatives. 

 \begin{algorithm}[H]
 \caption{ADP Generation}
 \begin{algorithmic}[1]
 \renewcommand{\algorithmicrequire}{\textbf{}}
 \renewcommand{\algorithmicensure}{\textbf{}}
 \REQUIRE
 \ENSURE  
  \textit{Initialisation}: \\
 $ADP_{i-1}$ = Last measurement \\
 $f'$ = Average first derivative in last 10 measurements  \\
 $f''$ = Average second derivative in last 10 measurements  \\
 $N_{ADP}$ = Number of ADPs required\\
  \FOR {$i = 1$ to $N_{ADP}$}
  \STATE $ADP_{i} = ADP_{i-1} + t_{s}f'$ \\
  $f' = f' + t_{s}f''$
  \ENDFOR
 \end{algorithmic}
 \end{algorithm}

\begin{figure}[H]
\centering
\captionsetup{justification=centering}
\includegraphics[width=7.5cm]{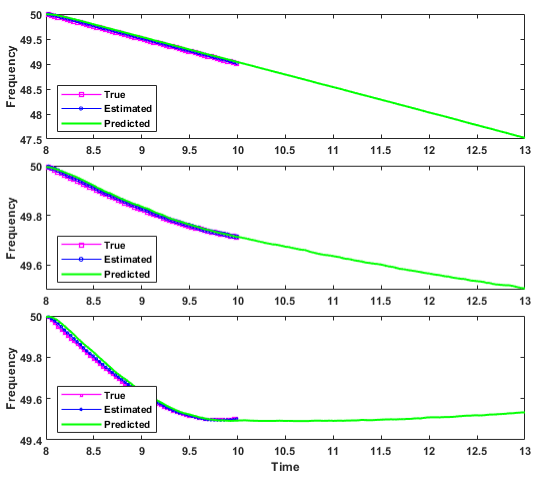}
\caption{PF predictions with varying degrees of curvature.\label{pic0}}
\end{figure}

The PF then processes the ADPs and iterates through predictions and correction steps to generate future estimates as illustrated in Figure \ref{pic0}.

The result is an algorithm that is able to capture the dynamics of the event using a single curve, unlike the models presented in \cite{larsson, bretas, rudez}. 

\subsection{Decision Making Problem}
After a prediction is made, equations derived from the swing equation are used to calculate the power imbalance \cite{relaybook}:

\begin{equation}
L = \frac{R_{p}H(1-\frac{f^{2}_{p}}{f^{2}_{1}})}{p(f_{p}-f{1})}\label{eq3}
\end{equation}

\begin{figure}[H]
\centering
\captionsetup{justification=centering}
\includegraphics[width=6.5cm]{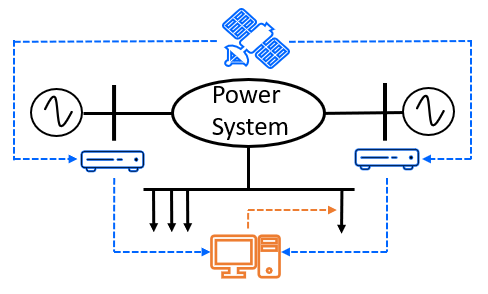}
\caption{Conceptual overview of the solution.\label{pic011}}
\end{figure}

$L$ represents the load excess factor, $H$ is the inertia factor, $p$ represents the power factor, and $R_{p}$ is the predicted average rate of change in frequency found with (4) using the current frequency measurement $f_{1}$, and the predicted frequency measurement $f_{p}$. 
In this work, frequency is predicted three seconds into the future; therefore, $f_{p}$ is the predicted frequency value three seconds after $f_{1}$. A visual overview of this framework is illustrated in Figure \ref{pic011}.

\section{Case Studies}

Three case studies are presented in this section. 
Each one utilizes the proposed algorithm to mitigate frequency deviations in slightly different ways. 
The simulations were performed in a reduced-order system model. 
This is an accepted assumption in the study power system frequency stability \cite{sakis}. 
All related works mentioned in Section I also made this assumption. 
In order to test the limits of this solution, a system with low inertia was chosen. 
In addition to the low inertia of the system, parameters in the speed controller of the generator were modified to decrease its performance. 
This produces a system where highly dynamic frequency deviations can be observed. 
These constraints would severely hinder the performance of the approaches suggested in \cite{larsson}, and in \cite{rudez}. 
The model used in the case studies and all its parameters can be found in \cite{sim}. 
Gaussian noise with a variance of 0.025 was added to the measurements.  

\subsection{Case Study I: Single Stage Load Shed}
\begin{figure}[H]
\centering
\captionsetup{justification=centering}
\includegraphics[width=8.5cm]{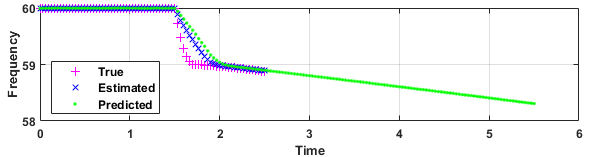}
\caption{Frequency deviation and prediction.\label{pic1}}
\end{figure}

In this scenario a single load shedding stage is used. 
Frequency deviations start at 1.5 s, as illustrated in Figure \ref{pic1}. Thresholds are exceeded and a prediction is made at 2.5 s.  
The frequency rate of change $R$ is calculated using (4). $f_{1}$ is the frequency value estimated at 2.5 s, while $f_{2}$ is the predicted frequency at 5.5 s. $R$ is then used in (7) to estimate the load excess factor $L$.  

\begin{figure}[H]
\centering
\captionsetup{justification=centering}
\includegraphics[width=8.5cm]{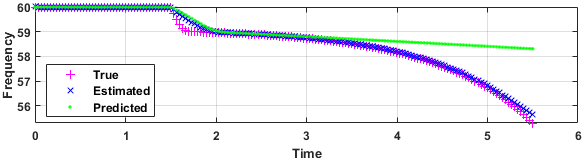}
\caption{Discrepancy between predictions and actual values.\label{pic2}}
\end{figure}

\begin{figure}[H]
\centering
\captionsetup{justification=centering}
\includegraphics[width=8.5cm]{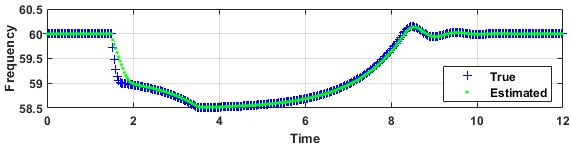}
\caption{Frequency at the end of the mitigation process.\label{pic3}}
\end{figure}

As depicted in Figure \ref{pic2}, the prediction in this case is not perfect; however, the algorithm still manages to bring the system frequency back to a level where generator governors can correct the deviation, as illustrated in Figure \ref{pic3}.

Corrective action is taken at 3.5 s, a full second after the thresholds were exceeded. 

The ability to take corrective actions early made up for a less than perfect prediction where the load excess factor was underestimated. The accuracy of the predictions and calculations can be improved by adjusting the thresholds and switching to a multi-stage scheme as shown in Case Study II.   

\subsection{Case Study II: Multi Stage Load Shed}
\begin{figure}[H]
\centering
\captionsetup{justification=centering}
\includegraphics[width=8.5cm]{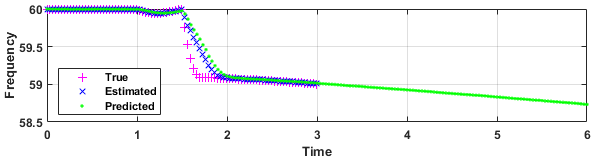}
\caption{Initial Prediction.\label{pic4}}
\end{figure}

In the second scenario, multiple load-shedding stages are used. This test case highlights the ability of the solution to adapt and compensate for inaccuracies made during the prediction step. In this case a prediction is made at 3 s, as shown in Figure \ref{pic4}.

\begin{figure}[H]
\centering
\captionsetup{justification=centering}
\includegraphics[width=8.5cm]{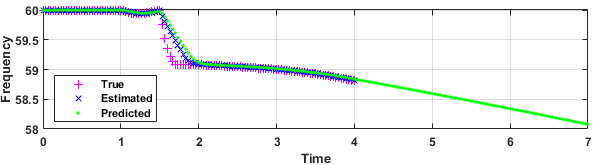}
\caption{Corrected prediction.\label{pic6}}
\end{figure}

\begin{figure}[H]
\centering
\captionsetup{justification=centering}
\includegraphics[width=8.5cm]{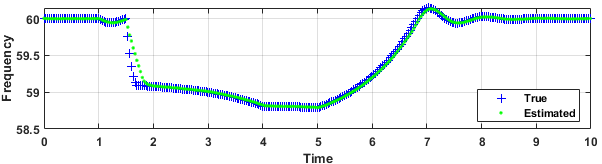}
\caption{Frequency at the end of the mitigation process.\label{pic7}}
\end{figure}

The same process seen in Case Study I is applied, except that this time only half of the calculated excess load  is dropped, while a new prediction is made one second after the first one.
\begin{figure}[H]
\centering
\captionsetup{justification=centering}
\includegraphics[width=8.5cm]{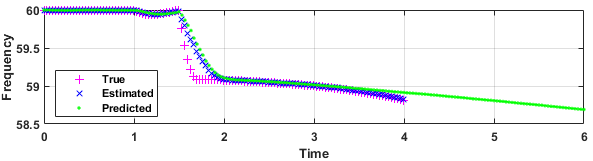}
\caption{Prediction and actual values.\label{pic5}}
\end{figure}
The first load shedding action takes place at 4 s. Meanwhile, the prediction is updated using data received after the initial prediction was made. This correction is illustrated in Figure \ref{pic6}.
With a new prediction, load-shedding action is once again executed at 5 s. Both load-shedding stages can be observed in Figure \ref{pic7}.

As illustrated in Figure \ref{pic7}, the first load-shedding stage stops the decline in frequency, while the second one sends it back to its normal range. A subsequent prediction is made at 5 s but no corrective action is taken as the algorithm predicts the frequency will be returning to normal levels.  

\subsection{Case Study III: Real Time UF Compensation with Distributed Energy Resources}
The goal of this final test case is to highlight the flexibility of the solution and use it to drive Distributed Energy Resources (DERs) in real-time. This test was run under a similar set of assumptions as those made by \cite{pul}. Once again, in order to test the algorithm under demanding conditions, faster frequency deviations than those seen in \cite{pul} were generated. Most importantly, the total delay time involved in the processing of data and actuation of DERs was increased to 500 ms; up from the 40 ms time delay used in \cite{pul}. That's a response time over ten times slower.

\begin{figure}[H]
\centering
\captionsetup{justification=centering}
\includegraphics[width=8cm]{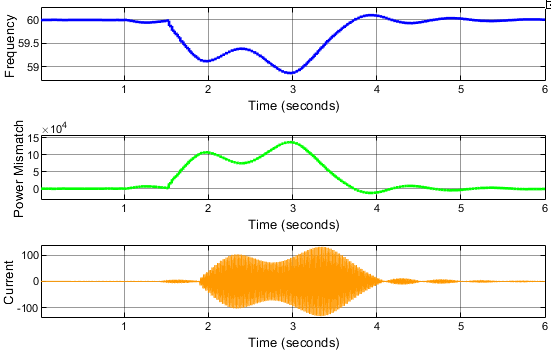}
\caption{Real-Time UF mitigation via DERs.\label{pic8}}
\end{figure}

Frequency deviations start at 1 s, with a significant loss in generation at 1.5 s. As illustrated in Figure \ref{pic8}, DER actuation take place 0.5 s after deviations in system frequency. 

As before, measurements are made via PMUs at 30 Hz. The power mismatch is calculated continuously in this test case per (4)-(7) in the form of a controller. As shown in Figure \ref{pic8} above, the frequency and power mismatch follow virtually the same trend but in opposite directions. When frequency deviates from the 60 Hz reference, a corresponding current output is seen from the DERs based on the power mismatch calculated. Despite a 0.5 second delay before DER actuation, the system successfully mitigates the frequency deviations.         


\section{Conclusion}
A proactive and automatic underfrequency load shedding solution was presented in this work. 
The solution leverages the particle filter along with existing technology to deliver a real-time and predictive UF mitigation scheme. 
Several limitations of existing techniques were addressed and improved upon, particularly in regards to real-life implementation. 
Test results indicate that regardless of the decision making strategy in place, load shedding or DERs actuation, frequency stability is considerably improved by this technique compared to traditional UF schemes. 
Results from the case studies also highlight the robustness and the flexibility of the particle filter. 




\vspace{12pt}

\end{document}